\documentclass[a4paper,11pt]{article}
\usepackage[T1]{fontenc} 
\RequirePackage{graphicx}
\usepackage{enumerate}
\usepackage{bm}
\usepackage{braket}
\usepackage{slashed}
\usepackage{comment}
\usepackage{color}
\title{Non-perturbative particle production and differential geometry}
\author{Tomohiro Matsuda\thanks{Laboratory of Physics, Saitama Institute
of Technology, Fukaya, Saitama 369-0293, Japan}}
\begin{document}

\maketitle

\abstract{
This paper proposes a basic method for understanding stationary
particle production on manifolds by means of the Stokes phenomenon.
We studied the Stokes phenomena of the Schwinger effect, the 
Unruh effect and Hawking radiation in detail focusing on the origin of
their continuous particle production. 
We found a possibility that conventional calculations may not explain
the experimental results.
}
\section{Introduction}
First, we will briefly summarise the motivation for this paper.
The Stokes phenomenon of the Schwinger effect is already well known,
 but it cannot be used naively as an explanation for stationary
particle production.
The main difficulty of this problem lies in the convention of using
 asymptotic states to define the vacuum.
To address the problem of the vacuum definition and stationary
particle production, we show how to understand the ``vacuum'' of the
Schwinger effect.
The Stokes phenomenon in the Unruh effect is then discussed and the
differences to the Schwinger effect are explained.
Normally, when calculating the Bogoliubov transformation of
the Unruh effect, it is derived from the consistency of the whole
space. 
However, if the constant acceleration is a temporal approximation, this
extrapolation may incorrectly imply distant correlations that do not
originally exist.
To answer this question, one has to derive an explicit local Unruh effect.
We have identified the Stokes phenomenon of the Unruh effect and
calculated the Bogoliubov transformation in local.
Our result (the Unruh temperature) differs from conventional
calculations by a factor of 2, 
which seems to be suggesting the absence of the distant correlation.
This will be tested by experimentation because this factor appears in
the Unruh temperature.
Then, we show that the Schwinger and the Unruh effects should occur
simultaneously under a strong electric field, which can be
 verified experimentally as an enhancement of the Schwinger effect.
Identifying the Stokes phenomena of particle production and
discriminating between similar phenomena will allow us to clearly
understand the origin of the phenomena.

The creation of particles from a vacuum has long been studied as a
fundamental problem in quantum mechanics\cite{Birrell:1982ix}.
Among them, the most fundamental scenario is the case when 
the mass of the particle varies with time\cite{Kofman:1997yn}.
The equation of motion of the particle is reduced to a second-order
ordinary differential equation, and particle production can be explained
by the appearance of the Stokes phenomenon, which mixes the two
solutions (a pair of plus and minus
sign solutions)\cite{Enomoto:2020xlf,Enomoto:2021hfv}.
In this case, the ``vacua'' are defined by two pairs of asymptotic
solutions, one in the past and one in the future, and particles
are created from the vacuum due to the different definitions of the
creation and the annihilation operators in the two limits\cite{Berry:1972na}.

Although the definition of the vacuum using asymptotic solutions looks
good, it is sometimes difficult to know how to define the vacuum.
For example, if one wants to see the Schwinger
effect\cite{Schwinger:1951nm, DiPiazza:2011tq} in a constant
electric field, one has
to assume states where the electric field artificially disappears in the
past and in the future, and furthermore the vacuum solutions have to be
adiabatically connected to the asymptotic states except for the region
where particles are created\cite{Haro:2010mx}.
If the constant electric field exists for a
long enough time in quantum terms, this approach seems quite artificial
 and leaves the simple question of why local analysis is not
 possible.

The situation is even more difficult in Hawking's original
paper\cite{Hawking:1975vcx}.\footnote{The theme of this paper is
``Solving stationary particle production by the local Stokes phenomena''.
There are many papers on Hawking radiation, but we will carefully
focus on our subject to avoid divergent explanations.}
Hawking calculates the Bogoliubov coefficient to estimate the radiation
from the black hole, using the natural vacuum before the gravitational
collapse as the in-vacuum, and the vacuum of the distant observer after
the black hole is created as the out-vacuum.
If Hawking radiation is produced by local physics near the horizon,
this analysis seems very artificial.
One reason for the need for such calculations is the
problem that the field equations in curved spacetime do not, by
mathematical definition, look at inertial systems by itself.
The mathematical definition starts with the tangent space, which is not
the inertial frame, but the Lorentz frame. 
The metric can be calculated from the vierbein, but the metric is
exactly the same for both the inertial frame and the Lorentz frame.
The only difference that appears is vierbein, which makes vierbein
essential for local calculations of Stokes phenomena.
This problem is also seen in the calculation of the Unruh effect.
Therefore, the Stokes phenomenon of the Schwinger effect is manifested
in the field equations, whereas such a Stokes phenomenon does not appear
in the field equations for the Unruh effect and Hawking radiation.
The Unruh effect and Hawking radiation require more than the field
equations to explain the Stokes phenomena.
Normally, it will be a global consistency condition or a collapse gap,
but we solved the Stokes phenomenon by focusing on the fact that the 
vierbein is the only way to view inertial systems locally.

Since these models are describing stationary radiations defined on
manifolds, it would be natural to assume that they can be explained in
 terms of the basic properties of manifolds.
Furthermore, if the solutions of the differential equations are given by
regular functions, these solutions do not mix until they cross the
discontinuity (the Stokes lines) on the complex time plane.
Taking this into account, it can be understood that in order to study
particle production, it is necessary to clarify the Stokes phenomenon at
the Stokes lines without using indirect methods.
Differential equations have independent solutions, but the relationship
between them and the vacuum solutions is not trivial. 
Usually the vacuum is defined somewhere (often asymptotically) and the
independent solutions are linked to the vacuum solutions there. 
The mixing of these solutions is allowed only at the discontinuity (the
Stokes lines).
Then, how can this ``vacuum definition'' be done locally on a manifold
for the stationary particle production?
These are the main motivations for this paper.

The Schwinger effect and Hawking radiation have also been analyzed using
path integrals\cite{Schwinger:1951nm, Hartle:1976tp},
 in addition to those based on field equations.
In the original paper\cite{Schwinger:1951nm}, Schwinger used a one-loop 
calculation to determine the ``vacuum decay rate'', noting that the
equation of motion does not result in a covariant formulation.
As will be explained later, the term ``vacuum'' used here should also be
treated with caution. 
The Schwinger effect is not the transition of the state $|0\rangle\rightarrow
|0'\rangle$, which must be accompanied by the domain walls.
The Schwinger effect should be explained in terms of the
Stokes phenomenon. We are not critical of Schwinger's calculations.
However, Schwinger's calculation about stationary particle production by
a constant electric field is indirect.
It is therefore natural to assume that new insights can be gained by
looking at the Stokes phenomenon directly, carefully incorpolating the
gauge degree of freedom to show why the particle production is
stationary.
If the electric field is explicitly time-dependent, the problem is
generally quite straightforward.
The reason is very simple.
This is because in models where the electric field is time-dependent,
the time of particle production is generally fixed at a particular time.
Our definition of vacuum aims to ``incorporate the gauge degrees of
freedom into the definition of vacuum'', but for the reasons given above,
a time-dependent electric field can be solved without
incorporating the gauge degrees of freedom.
In this case, defining the vacuum asymptotically or in tangent space
makes no difference.
Conversely, this was the cause of the postponement of the solution of
the steady-state radiation problem.
See Schwinger's series of papers\cite{Schwinger:1951xkv1,
Schwinger:1953tbv2, Schwinger:1953zzav3, Schwinger:1953zzv4,
Schwinger:1954zzav5, Schwinger:1954zzv6} and Ref.\cite{DiPiazza:2011tq}
for many issues not addressed in this paper.

About the Schwinger effect, it is worth noting that when Bogoliubov
coefficients of a charged scalar field are calculated 
using the Stokes phenomena, exact solutions are given by parabolic
cylinder functions\cite{Birrell:1982ix,Haro:2010mx,Dumlu:2010ua}.
These exact solutions are solutions in the presence of an electric
field.
There is a problem here. While the solution of the equation
of motion indicates that particle production occurs at a specific time,
this time should be arbitrary given the degrees of freedom of the
gauge. How can the equation of motion and the degrees of freedom of the
gauge successfully coexist? We turn to the definition of the vacuum.
If a constant electric field of the Schwinger effect 
exists almost forever and the radiation is stationary, a very simple
question arises: where exactly is the local ``vacuum'' on the manifold?

Our answer to this question is as follows.
Given the differential geometry of the manifold, a tangent vector
space can be defined at any point.
Then, we define the vacuum using the tangent vector.
It is worth elaborating on this definition.

First, we describe the manifolds with local Lorentz transformation.
Mathematically, due to Lorentz symmetry, there are an infinite number of
ways to define the vacuum, so one might think that the phrase
``defining the vacuum using the tangent vector'' leaves an ambiguity.
Of course, the vacuum is gauge and Lorentz invariant, but
the field equations do not explicitly implement these degrees of
freedom.
Therefore, the question is ``how to define the vacuum for the practical
calculation of the Stokes phenomena usign equations of motion,
incorpolating the degrees of freedom''.
The use of the equations of motion is essential for the study of the
Stokes phenomenon, which is why this paper is devoted to this subject.
This can be resolved by considering the difference between the treatment
of manifolds as mathematics and the treatment of manifolds in the
description of physical phenomena.
Mathematical manifolds are constructed so that they are consistent for
all observers. 
On the other hand, when discussing physical phenomena, there are
observers, so the observer's frame of reference is naturally chosen.
So, who is the ``observer'' in the Schwinger effect?
Opinions may differ as to whether the person performing the experiment
should be the ``observer'' or whether the generated particles are the
``observers'' of the vacua, but to understand the mechanism of the
Schwinger effect one 
needs to understand the vacuum for the generated particles we are
looking at. 
See also the argument in Ref.\cite{DiPiazza:2011tq}.
Each particle produced has a different momentum, and the ``vacuum'' for
the particle is defined by its own unique (rest) frame. 
Although it may seem that the gauge symmetry is important for the
Schwinger effect and the Lorentz transformation is irrelevant, actual
calculations of the Stokes phenomenon show that the Lorentz
transformation plays an important role in understanding stationary
radiation.
Further explanation will be given later, as specific
differential equations need to be given for the explanation.

If we consider gauge transformations on manifolds, what is the
definition of a vacuum?
In the case of Lorentz symmetry, the observer's frame has to be 
chosen for the vacuum.
In this way, there was no speed gap between the observer and the
frame.
Similarly, we choose the vacuum on $A_\mu=0$ section,\footnote{ 
Some discussion about the incompatibility between the gauge $A_\mu=0$
and the path integrals is 
explained on page 295 of the textbook\cite{Peskin-textbook}
by Peskin and Schroeder.} 
taking into account the continuity of the tangent vectors and the
covariant derivatives.
The above vacuum choices may seem trivial, but they are crucial to
understanding the Stokes phenomenon of the steady-state radiation on
manifolds  using equations of motion.

Defining the vacuum in this way avoids the need to artificially
introduce asymptotic states.
Such asymptotic states might have been essential in analyses using the
usual WKB approximation, but thanks to the development of the Exact
WKB(EWKB)\cite{RPN:2017,Delabaere:1993,Silverstone:2008, Voros:1983,
Virtual:2015HKT,
WKB-recent:2022a,WKB-recent:2022b,WKB-recent:2022c,WKB-recent:2022d,WKB-recent:2022e,WKB-recent:2022f,WKB-recent:2022g,WKB-recent:2022h,WKB-recent:2022i}, 
 the local structure of the Stokes phenomenon is now sufficiently well
 understood in terms of resurgence.
In this paper, we will discuss what happens if stationary particle production 
is solved by means of the local Stokes phenomena.

With regard to the definition of a vacuum, it is also shown that 
the vacuum of Hawking radiation must be defined on
a local inertial frame, not on the local Lorentz frame.
The crucial difference between the two frames is not very clear
in many textbooks, so this paper uses a vierbein to illustrate the
difference.
Suppose we write the equations of motion for an
accelerating observer of the Unruh effect.
Let us first consider the tangent space (Lorentz) and introduce covariant
derivatives using the metric.
When calculating the metric, the inertial system vierbein might be used,
but this metric is identical to the Lorenz frame.
Eventually, traces of inertial systems disappear from the equations of
motion.
Unlike the Schwinger effect, this makes it impossible in principle to
derive the Stokes phenomenon of the Unruh effect directly from the
equation of motion.
This is the reason why the Unruh effect required a global analysis.
We present a local analysis of the Unruh effect
by using the vierbein, the only local clue to the inertial system.

The idea of defining a vacuum in an inertial system and calculating
Bogoliubov coefficients for an accelerating observer is beautifully
summarized in Ref.\cite{Birrell:1982ix} for the Unruh
effect\cite{Unruh:1976db}.
However, their analysis uses the global nature of the Rindler
coordinates and this may introduce unphysical correlations.
Indeed, if the solution with constant acceleration is a 
local (and a temporal)
approximation, the results obtained by extrapolating it to the whole
space are questionable.
In particular, the strong correlation between two distant wedges is
likely to be a by-product of this extrapolation.
The local analysis in this paper produces a different result (the Unruh
temperature) from the global analysis.
We refer to this difference as the factor 2 problem.
We believe that it is the experimental verification\cite{Bell:1982qr}
 of this factor 2
problem that is most important for understanding the Unruh effect.

First, we discuss the Schwinger effect in a constant electric field as an
obvious example in gauge theory.

Then we discuss Hawking radiation and the Unruh effect\cite{Unruh:1976db}.
In contrast to the Schwinger effect, it is not the connection that is
important for the local Stokes phenomenon:
the Stokes phenomenon cannot be seen in the ``same'' way as the
Schwinger effect, as we have described above.

Although the Schwinger effect and Hawking radiation share many
similarities, these comparisons using differential geometry and the
Stokes phenomenon highlight crucial differences between them.

\section{The Schwinger effect}
\begin{figure}[h]
\centering
\includegraphics[width=0.8\columnwidth]{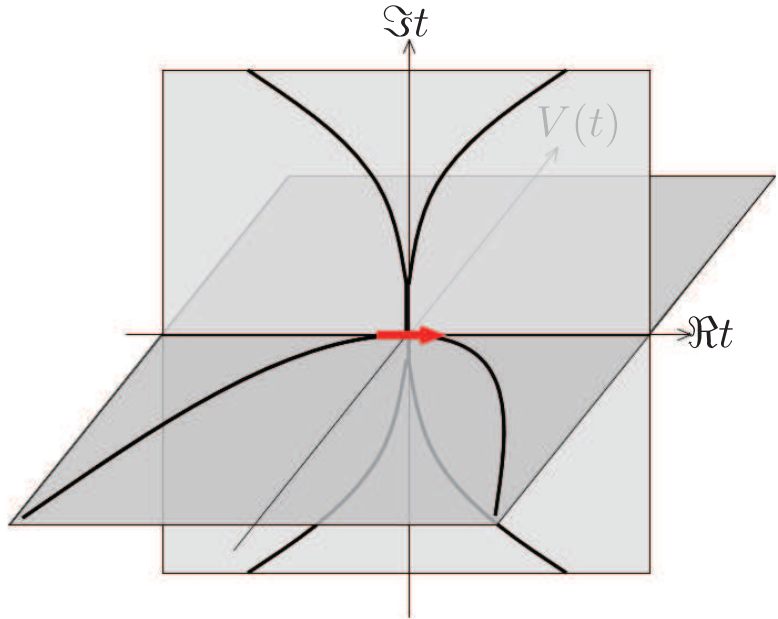}
 \caption{The potential $V(t)=-q^2E^2t^2$ and the Stokes lines (on the
 complex $t$ plane) are
 shown. Particle production occurs when the real $t$-axis crosses the
 Stokes line (indicated by the arrow at the origin).}
\label{fig-1}
\end{figure}

Here we consider the case where the electric field is
spatially homogeneous and has a constant value in the z-direction.
For a complex scalar field $\phi$ of mass $m$ in a four-dimensional
Minkowski spacetime, we consider the action $S_0$ on a tangent space given
by 
\begin{eqnarray}
S_0&=&\int d^4x \left(\partial_\mu \phi \partial^\mu\phi^*-m^2
	       \phi\phi^*\right).
\end{eqnarray}
Introducing a gauge field $A_\mu$, the partial derivatives are replaced
by covariant derivatives of the differential geometry:
\begin{eqnarray}
\partial_\mu &\rightarrow& \nabla_\mu \equiv\partial_\mu+i q A_\mu.
\end{eqnarray}
We define the vacuum on the tangent space attached to $A_\mu=0$.
Assuming the limit where dynamics of the gauge field is negligible (the
gauge field does not propagate in this limit), the 
gauge field is external and given by
\begin{eqnarray}
A^\mu&=&(0,0,0, -E (t-t_0))
\end{eqnarray}
with the electric field strength $\vec{E}=(0,0,E)$ for an arbitrary
$t_0$.
For the scalar field $\phi$, the equation of motion after Fourier
transformation is
\begin{eqnarray}
\label{eq-field-Sch}
\ddot{\phi}_k+\omega_k^2(t)\phi_k&=&0,
\end{eqnarray}
where $\omega_k^2(t)=m^2+k_\perp^2+(k_z-qE(t-t_0))^2$.
The exact solutions are described by using the parabolic cylinder
functions\cite{Birrell:1982ix,Kofman:1997yn} or the Weber
functions \cite{Enomoto:2020xlf}.
Using the EWKB, one can find a Merged pair of Turning Points
(MTP)\cite{Enomoto:2020xlf} whose Stokes line crosses on the real axis at
$t=t_0$.
The stokes lines are shown in Fig.\ref{fig-1}.

To describe the Bogoliubov transformation explicitly, we expand
$\phi_k(t)$ using the solutions $\psi^\pm_k(t)$ of
Eq.(\ref{eq-field-Sch}) as 
\begin{eqnarray}
\label{eq-WKB}
\phi_k(t)&=& \alpha_k \psi^-_k(t)
+\beta_k\psi^+_k(t).
\end{eqnarray}
The transformation matrix is then given by
\begin{eqnarray}
\left(
\begin{array}{c}
\alpha_k^R\\
\beta_k^R
\end{array}
\right)
&=&
\left(
\begin{array}{cc}
\sqrt{1+e^{-2\pi \kappa}}e^{i\theta_1} & ie^{-\pi\kappa+i\theta_2}\\
-ie^{-\pi\kappa-i\theta_2} &\sqrt{1+e^{-2\pi \kappa}}e^{-i\theta_1} 
\end{array}
\right)
\left(
\begin{array}{c}
\alpha_k^L\\
\beta_k^L
\end{array}
\right),\nonumber\\
\end{eqnarray}
where L (On the left hand side of the Stokes line) and R (On the right
hand side of the Stokes line) are for $t<t_0$ and $t>t_0$, respectively.
The constant $\kappa$ is calculated from the integral connecting the two
turning points $t_*^\pm$ appearing on the imaginary axis and
it becomes $\kappa=\frac{m^2+k_\perp^2}{2E}$.
Here, all the phase parameters are included in $\theta_{1,2}(k)$.
It is well known\cite{Haro:2010mx} that Schwinger's result (vacuum decay rate) 
can be reproduced by using this result and adding up for all
possibilities.
The only remaining problem is that the above equation describes particle
production at a specific time while this time ($t_0$) should be arbitrary
given the gauge degrees of freedom.
It is therefore necessary to consider specific ways of incorporating the
degrees of freedom of the gauge into the equations of motion.
The problem is that the conventional definition of vacuum, which uses
asymptotic states, does not explain the situation well.

\begin{figure}[h]
\centering
\includegraphics[width=0.5\columnwidth]{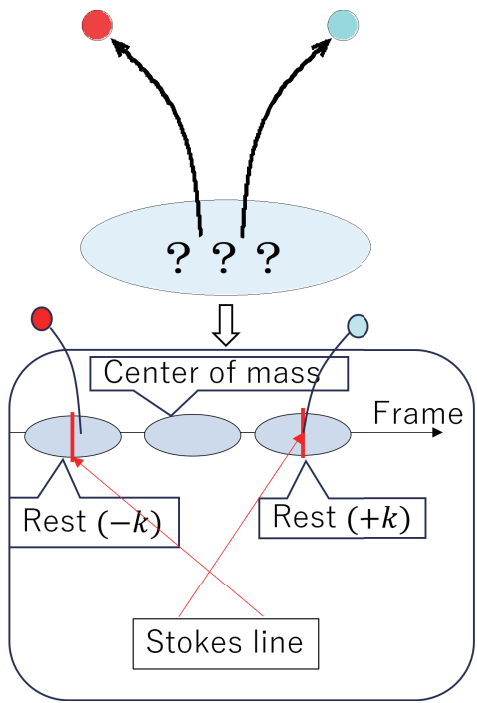}
 \caption{This figure shows the relationship between the Stokes lines and
 the frames when $k_z\ne 0$ in the center of mass frame. }
\label{fig-pmk}
\end{figure}

Now we solve the equation with $\bf{\it k}$ set to 0 (or $\bf{\it k}^2\ll
\omega^2$), since we are choosing the rest frame of the
particle.
Note that $k_z\ne 0$ shifts the position
of the Stokes lines. 
As we believe that the stationary particle production is realised
when the Stokes lines and the vacuum always coincide, choice of the
frame is very important.
Here, the frame is determined by naturalness and requests for stationary
particle generation.
See Ref.\cite{DiPiazza:2011tq} for other discussions.
Fig.\ref{fig-pmk} shows the displacement of the Stokes line (center of
mass frame) and how it coincides with the defined vacuum in the rest
mass frames.
We omit explicit Lorentz factors of $E$ and $\omega$ just for simplicity.
Eq.(\ref{eq-field-Sch}) looks like the Schr\"odinger equation and can be
solved as a scattering problem for the Schr\"odinger equation with the 
potential given by
\begin{eqnarray}
V(t)&=&-q^2A_z^2=-q^2E^2t^2,
\end{eqnarray}
where we set $t_0=0$ for simplicity.
The potential is shown in Fig.\ref{fig-1} together with the Stokes line on the
complex $t$-plane.

\begin{figure}[h]
\centering
\includegraphics[width=0.5\columnwidth]{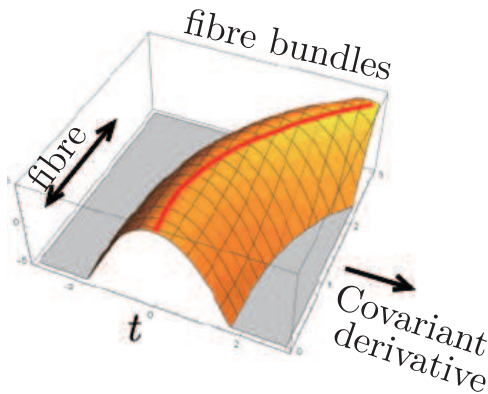}
 \caption{The potential ($V(t)=-q^2E^2t^2$) is illustrated on
 the manifold taking into  account the gauge degrees of freedom.
Since the Stokes phenomenon occurs on the vacuum attached to 
the red line (the line connecting vertices of the potential), the Stokes
 phenomenon is always found.}
\label{fig-2}
\end{figure}

Let us now define the vacuum not as an asymptotic state but in the
tangent space at $A_\mu=0$.
In this simplest situation, choosing $A_\mu=0$ gauge, 
one can ``always'' find a local vacuum
attached to $A_\mu=0$ on which the vacuum and the Stokes line coincide.
This is the very reason why particle production of the Schwinger effect can be
stationary.
The situation is illustrated in Fig\ref{fig-2}.
The Stokes phenomena are therefore properly defined without
relying on artificial asymptotic states, and the
definition of the ``vacuum'' is consistent with stationary radiation.
The freedom of the gauge is thus properly incorpolated.
The importance of the frame should be mentioned here.
If the frame had not been chosen to be the rest frame of the particle,
the Stokes line would not appear on the vacuum at $A_\mu=0$ due 
to the fact that $k_z\ne0$. 
(See the definition of $\omega^2_k(t)$.)
This may seem strange, but if the Stokes phenomena are
considered for the centre of mass frame, the Stokes line does not appear
just at the vacuum. 
The Stokes lines must be considered for the
respective rest frames for particles and antiparticles.

We have seen that on manifolds one can always find the vacuum of
particle creation which is defined locally on tangent space by choosing
the section $A_\mu=0$ and the rest frame of the particle.
So far, our calculation is fully consistent with the Schwinger's
original calculation, but after calculating the local Unruh effect we
will show that we have to introduce a difference.

We have seen that the Stokes lines always coincide with the defined vacuum.
For the Schwinger effect, we have seen that the local Stokes
phenomenon is induced simply by the covariant derivatives.
Let us now show that this is not the case with Hawking radiation
and the Unruh effect.
The cause of the failure tells us where the Stokes
phenomenon comes from in the Unruh effect.

\section{The Unruh effect and Hawking radiation}
Following the Schwinger effect, one can define a local vacuum
to find stationary radiation due to the Stokes phenomenon.
The vacuum of the particle production is defined by choosing a frame for which the
particle is at rest.
In general relativity, there are two different definitions of coordinate
systems in which the above condition is satisfied: the (local) Lorentz
frame and the local inertial frame.
Although the two frames are often considered to be physically almost
identical because they give the same metric, the difference between the
two frames is crucial when discussing stationary radiation on manifolds.

In order to understand the difference with gauge theory, it is first
necessary to understand what happens if the derivatives are naively
replaced by covariant derivatives.
This manipulation gives the Klein-Gordon equation of an accelerating
observer (Unruh) or on curved spacetime (Hawking), but unlike the
Schwinger effect, it cannot explain the 
local Stokes phenomenon\cite{Enomoto:2022mti}.\footnote{Stokes
events around black holes can occur outside the horizon.
We discuss Stokes events that are directly related to local Hawking
radiation at the event horizon.
See Ref.\cite{Enomoto:2022mti,Dumlu:2020wvd} for the Stokes
phenomena appearing outside the horizon and their
contribution to the glaybody factor.}
As already explained, the root of this problem is due to the fact that
the equations of motion of the accelerating system are based on the
Lorenz system and do not look at the inertial system.
The only difference can appear in the vierbein, but
there was no analysis of the Stokes phenomenon using this until
Ref.\cite{Enomoto:2022mti}.
The Bogoliubov transformation can be computed if the solution is
extrapolated and analysed in the entire space, but there is a risk that
extra information could be introduced by the extrapolation.
We speculate that strong correlation between distant wedges is this
``extra information''.
We refer to this problem as the factor 2 problem.

To understand the essence of the story, it is better to consider the
Unruh effect before Hawking radiation.
The vacuum seen by an observer of the Unruh effect is the inertial
system, not the Lorentz system. 
This can be confirmed by computing the vierbein.
On the other hand, what the equation of motion of the
accelerating observer can see is the Lorentz system (by definition).
The metric is exactly the same whether it is a local inertial
system or a Lorentz system, so we rarely distinguish between the
two, but in the present case there must be some crucial difference.
In fact, when calculating the vierbein of the Rindler spacetime given by
\begin{eqnarray}
\label{eq-rindler-t}
t&=&\frac{1+\alpha x_r}{\alpha}\sinh (\alpha t_r)\nonumber\\
x&=&\frac{1+\alpha x_r}{\alpha}\cosh (\alpha t_r),
\end{eqnarray}
which describes the coordinate system of an object
moving at constant acceleration $\alpha$ through a flat space-time
represented by $(t,x)$, one will find
\begin{eqnarray}
dt&=&\left(1+\alpha x_r\right)\cosh (\alpha t_r) dt_r
+\sinh(\alpha t)dx_r\nonumber\\
dx&=&\cosh(\alpha t)dx_r +\left(1+\alpha x_r\right)\sinh (\alpha t_r) dt_r.
\end{eqnarray} 
Obviously, the vierbein is looking at the inertial system.
The metric is calculated as
\begin{eqnarray}
g_{\mu\nu}&=&\eta_{mn}e^m_\mu e^n_\nu,
\end{eqnarray}
where $\eta_{mn}$ is for the local Minkowski space.
This gives the Rindler metric given by 
\begin{eqnarray}
\label{eq-metric-R}
ds^2&=&-(1+ \alpha x_r)^2dt_r^2+dx_r^2.
\end{eqnarray}

Note that the metric is identical for both (inertial and the Lorentz)
frames but the explicit $t_r$-dependence of the vierbein appears only
for the inertial system.
(The vierbein of the Lorentz frame is diagonal.)
As we will see, the $t_r$-dependence of the vierbein is where the Stokes phenomenon of stationary
radiation on curved space-time comes from.

As is summarized in Ref.\cite{Birrell:1982ix}, it is possible to
calculate the Bogoliubov coefficients by considering carefully 
the global structure of the Rindler coordinates and the relationship
between the vacuum solutions written in the two coordinate systems.
Let us elaborate on this (global) calculation first so that the reason
for the distant correlation can be clearly understood.

We introduce the Rindler coordinates $(\tau,\xi)$ in the right wedge as
\begin{eqnarray}
t&=&\frac{e^{\alpha\xi}}{\alpha}\sinh a\tau\nonumber\\
z&=&\frac{e^{\alpha\xi}}{\alpha}\cosh a\tau,
\end{eqnarray}
where $(t,z)$ are the coordinates of the Minkowski space.
Introducing the light-cone coordinate system of the Rindler space as
$u=\tau-\xi, v=\tau+\xi$, we find that they are connected to the
light-cone coordinate system of the Minkowski space $(U,V)$ as
\begin{eqnarray}
U&=&-\frac{e^{-\alpha u}}{\alpha}\nonumber\\
V&=&\frac{e^{\alpha v}}{\alpha}.
\end{eqnarray}
In the left wedge, we define the Rindler coordinates
$(\tilde{\tau},\tilde{\xi})$.

Just for simplicity we assume massless particles and consider only the
right-moving waves.
In the inertial system, the field $\phi(U)$ can be expanded as
\begin{eqnarray}
\phi(U)&=&\int^\infty_0dk \left[a_k f_k^{(M)}(U)+a_k^\dagger
			   f_k^{(M)*}(U)\right],\nonumber\\
f_k^{(M)}(U)&=&\frac{e^{-ikU}}{\sqrt{4\pi k}},
\end{eqnarray}
which defines the inertial vacuum $a_k|0_M\rangle=0$

For the right Rindler system $(\tau,\xi)$, we have
\begin{eqnarray}
\phi_R(u)&=&\int^\infty_0dp \left[b_p^{(R)} f_p^{(R)}(u)+b_p^{(R)\dagger}
			   f_p^{(R)*}(u)\right],\nonumber\\
f_p^{(R)}(u)&=&\frac{e^{-ipu}}{\sqrt{4\pi p}}=\theta(-U)\frac{(-\alpha
 U)^{ip/\alpha}}{\sqrt{4\pi p}},
\end{eqnarray}
which defines the Rindler (right) vacuum $b_p^{(R)}|0_R\rangle=0$,
and same calculation defines the Rindler (left) vacuum.

Although the phase space at $t=0$ can be given by
$|M\rangle=|L\rangle\otimes|R\rangle$, it does not mean that the vacuum
satisfies $|0_M\rangle=|0_L\rangle\otimes|0_R\rangle$.
Indeed, using the above solutions one can find
\begin{eqnarray}
\label{eq-def-vac-unruh}
|0_M\rangle&\propto&\exp\left[-\prod_p 
\left(e^{-\pi p/2\alpha}b_p^{L\dagger}\right)
\left(e^{-\pi p/2\alpha}b_p^{R\dagger}\right)\right]|0_L\rangle\otimes|0_R\rangle,
\end{eqnarray}
which shows strong correlation between distant wedges.
After normalization and trace about the left Rindler states, one will
find the Unruh temperature $T_U=\hbar \alpha/2\pi c k_B$. 
What is important here is the duplication of the factor 
$e^{-\pi p/2\alpha}$ due to the correlation between distant wedges.
This is the source of our factor 2 problem.

The global calculation ``revealed'' a surprising
correlation of particle production between two causally disconnected
wedges.
More details and different approaches can be found in Ref.\cite{Birrell:1982ix}.
However, it may seem unnatural that global information is essential when
its motion can be viewed as constant acceleration only over a certain period of time.
Here, what we want to consider is a local system defined
in the neighborhood of a point.

Let us see how the Stokes phenomenon of the Unruh effect appears
locally.
During the Unruh effect, an accelerating observer is looking at the
inertial vacuum.
Therefore the vacuum solutions of the inertial system have to be seen by
an observer using the vierbein.
To keep the integral of the solutions intact, we use
$dt=\cosh (\alpha t_r) dt_r$
to write the vacuum solutions after the Fourier transform
into
\begin{eqnarray}
\label{eq-rindler-sol}
\phi_k^\pm(t)&=&A_k e^{\pm i \int \omega dt}\nonumber\\
&=&A_k e^{\pm i \int \omega_k \cosh(\alpha t_r) dt_r}.
\end{eqnarray}
We are choosing the particle's rest frame, for which ${\bf \it k}^2 \ll \omega^2$. 
It is normally difficult to recognize the Stokes phenomenon from these
solutions, but the problem can be solved using the basic properties of
the EWKB.
First define $Q(t)_0\equiv -\omega_k^2\cosh^2(\alpha t_r)$ and consider
 the following ``Schr\"odinger'' equation
\begin{eqnarray}
\label{eq-rindler-EoM}
\left(-\frac{d^2}{dt^2}+\eta^2 Q(t,\eta)\right)\psi(t,\eta)&=&0,
\end{eqnarray}
where $\eta\gg 1$ and $Q(t,\eta)$ is expanded as
\begin{eqnarray}
Q(t,\eta)&=&Q_0(t)+\eta^{-1}Q_1(t)+\eta^{-2}Q(t)+\cdots.
\end{eqnarray}
The solution of this equation can be written as $\psi(t,\eta)\equiv
e^{\int S(t,\eta) dt}$, where $S(t,\eta)$ can be expanded as 
\begin{eqnarray}
S&=&S_{-1}(t)\eta +S_0(t)+S_1(t)\eta^{-1}+\cdots.
\end{eqnarray}
The point of this argument is that after introducing $\eta$ properly
in Eq.(\ref{eq-rindler-sol}), one can choose $Q_i(t), i\ge 1$ to
find the ``Schr\"odinger equation'' which gives the solutions 
Eq.(\ref{eq-rindler-sol}).\footnote{See also Section 5 of
Ref.\cite{Costin:2008}.}
Here $\hbar$ of quantum mechanics has been replaced by $\eta$ according to
mathematical convention.

\begin{figure}[h]
\centering
\includegraphics[width=0.8\columnwidth]{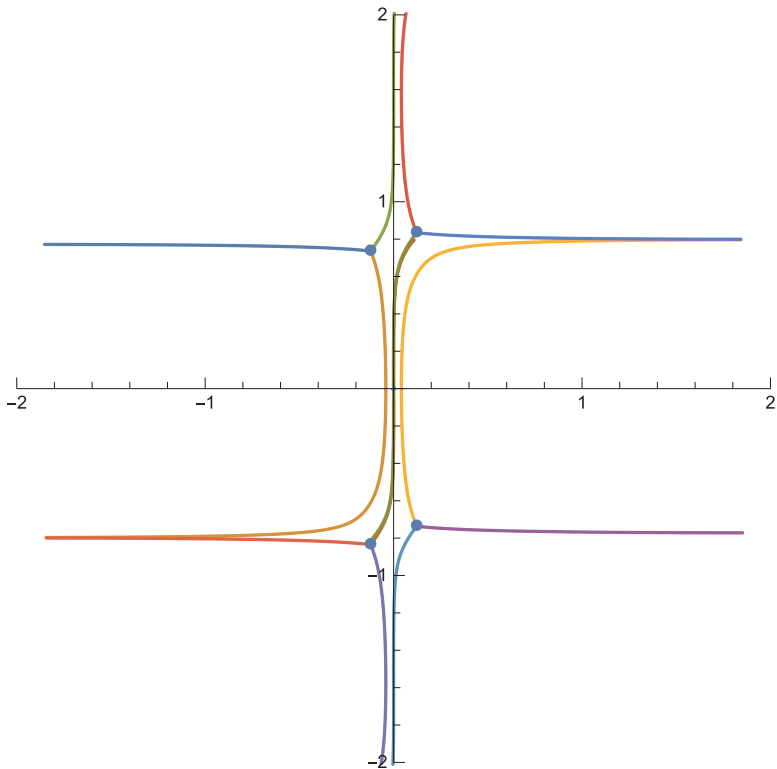}
 \caption{The Stokes lines of the Unruh effect is shown for
 $Q(t)=-(Cosh^2 (2t)+0.05+0.05 i)(1-0.05i)$. The degenerated Stokes
 lines are separated introducing small parameters.}
\label{fig-Unruh-stokes}
\end{figure}

This procedure allows us to make use of a powerful analysis of the
EWKB: one can calculate the Stokes lines only by using 
$Q_0(t)$.
After drawing the Stokes lines, one can see that a Stokes line crosses on the
real axis at the origin\cite{Enomoto:2022mti}.
The Stokes lines of the Unruh effect is shown in
Fig.\ref{fig-Unruh-stokes}.
This allowed us to expand $Q(t)_0$ near the origin and finally we have
\begin{eqnarray}
Q(t)_0&=& -\omega_k^2\cosh(\alpha t_r)\nonumber\\
&\simeq&-\omega_k^2-\alpha^2 \omega_k^2 t_r^2,
\end{eqnarray}
which gives a familiar Schr\"odinger equation of scattering by an
inverted quadratic potential.
The equation can be solved using the parabolic cylinder functions or the
Weber functions, giving the MTP structure of the Stokes
lines\cite{Enomoto:2020xlf,Enomoto:2021hfv,
Enomoto:2022mti}.\footnote{See also Fig.\ref{fig-1}.}
The Stokes phenomenon occurs at $t=0$, which can be regarded as the
point of contact ($v=0$) between the observer in the accelerating system
and the vacuum in the inertial system. 
Again, the Stokes lines coincide with the vacuum only when
the vacuum is defined for the rest frame of the particles, not for the
experimenter.
This choice of the frame is of course consistent with other
analysis\cite{DiPiazza:2011tq}.

The Bogoliubov coefficients of this calculation generate the Boltzmann
factor $\sim e^{-\pi \omega_k/\alpha}$.
As we have already mentioned, the global analysis of Ref.\cite{Birrell:1982ix} 
gives $\sim e^{-2\pi \omega_k/\alpha}$, which gives the probability of
strongly correlated particle production in the two distinct regions and
there is the factor 2 problem.
The duplication of the factor occurs due to the correlation.
We can see that two particles are produced simultaneously in global
analysis (which uses extrapolation) while we are calculating local and
single particle production. 
Whether this correlation actually exists can be
verified by experimentation using our results, since it appears simply
in the Unruh temperature.
The same applies to Hawking radiation, because Hawking radiation
requires creation of a pair of negative and positive energy
particles across the horizon, but only the positive energy
particle outside the horizon can be observed as radiation.
This is the typical ``2 for 1'' particle production.
Two particles are produced but only one can be observed.
In this case, the production probability of one particle ($P_1$) should
be discriminated from the observation probability of one particle
($P_1^{obs}$). 
Therefore, our local analysis distinguishes ``one particle production
rate $P_1$'' from ``one particle observation rate $P_1^{obs}(=$
two particle production rate $(P_1)^2$)''. 
There is no factor 2 problem in Hawking radiation because the ``2 for
1'' particle production is essential due to the conservation of the
energy law.

\section{Simultaneous Schwinger-Unruh-Hawking effect}
Finally, it should be mentioned that the Schwinger and the Unruh effects
(Hawking radiation without the event horizon because the strong electric
field allows pair creation) occur simultaneously under
strong electric fields.\footnote{See also 
Ref.\cite{Ritus:2002rq,Ritus:2003wu,Ritus:2015qnt}, in
which similarities and differences between the Schwinger and the Unruh
effects are discussed.
The crucial difference with this paper is whether it deals with the
local Stokes phenomenon for the stationary particle production.
See also \cite{Volovik:2022cqk}, although the understanding of the
principle of the Unruh effect is completely different from this paper
and therefore synergies with the Schwinger effect have been analysed in a 
different way.
As already mentioned, we believe that identification of the Stokes
phenomenon leads to mathematical understanding of the physical
phenomena.}
As noted above, the way to choose the vacuum for stationary particle
production is ``look at local inertial
system, choosing rest frame for the particle and the gauge $A_\mu=0$''.
Considering that the  generated particles are
accelerated by the electric field, the particles are generated from 
``the vacuum seen by an accelerating observer (particle)''.
This is the same situation as in the case of the Unruh effect.
Furthermore, unlike the Unruh effect, pair production is possible
without violating the law of conservation of energy if the electric
field is strong enough.
Therefore, particle production is observed even in the absence of the event
horizon. 
Furthermore, since one of the two particles of a pair 
does not disappear into the horizon, both are observed.
As a result, this is not a ``2 for 1'' particle production and
the Unruh temperature observed simultaneously with the
Schwinger effect should show a factor of 2 difference compared to the
conventional (global) Unruh effect.

Perhaps the most interesting aspect of this story is whether it is
possible to verify experimentally that the Schwinger and Unruh effects
occur simultaneously.
As shown in this paper, the two effects arise from separate physical
phenomena, and  might or might not contribute in the same form.
When written in the EWKB notation, it might appear that there 
are two contributions $Q_0^{Schwinger}(t)$ and $Q_0^{Unruh}(t)$ to
$Q_0(t)=Q_0^{Schwinger}(t)+Q_0^{Unruh}(t)$, 
so that the coefficient of the inverted quadratic potential
causing the Stokes phenomenon is the sum of the two contributions. 
One issue arises here.
From the calculations so far, it is not surprising that the
Schwinger and Unruh effects can occur as independent phenomena. 
In that case, the result should be
$\beta^{tot}=\beta^{Unruh}+\beta^{Schwinger}$.
Which is correct can be determined by experimental verification.
The acceleration of a charged particle in a strong electromagnetic field
is $\alpha=qE/m$, which, when used in ``our'' equation for
the Unruh effect, 
gives an inverted quadratic potential with exactly the same coefficient as
the Schwinger effect.
If our considerations are correct, the actual probability of particle
production should be increased, and the quantitative property is
significantly different from Schwinger's calculation.
In this way, the simultaneous occurrence of the Unruh and Schwinger
effects is a phenomenon that can be demonstrated experimentally.
The previously mentioned problem with factor 2 of the Unruh effect 
can also be confirmed by experimenting.
If, on the other hand, the amplification noted here did not occur, then
it can be concluded that the Schwinger and Unruh effects are in fact
identical physical phenomena that are linked even more deeply in the
theory.

\section{Conclusions and discussions}

Particle production from a vacuum, including the Schwinger effect, Hawking
radiation and the reheating of the universe, is a very important area of
research in theoretical physics.
We believe that identification of the Stokes phenomena is essential to
understand these phenomena. 
However, conventional WKB approximation does not allow such 
analysis at locations where the adiabatic condition is violated.
The development of the EWKB largely solved this problem, but the
practice persisted for a long time, as important papers of the time set
up asymptotic vacuum at a far distance.
Serious problem does not arise in simple models because in such models
only the physical quantity (e.g., the mass) changes with time.
However, in cases where the physical quantity does not change, such as
the Schwinger effect in a constant electric field, the Stokes phenomenon
of the solutions raises major questions about the definition of the
vacuum.
This paper aims to establish the definition of the vacuum in
such physical phenomena using the notion of manifolds to identify the
Stokes phenomena.

The benefits of our analysis are significant:
without defining a vacuum at a distance or creating a collapse gap,
local stationary radiation can be analyzed using manifolds.
Our results differed from the conventional calculation of
the Unruh effect by a factor of 2.
This factor is an indicator of whether the correlations that appear in
traditional calculations are correct.
This paper also discusses the possibility that the Unruh effect may
alter the results of the Schwinger effect.
All of these can be demonstrated by experimentation.

It is the development of the EWKB that makes it possible to deal with the
seemingly complex equations that result from such analysis.
By understanding the local nature of the Stokes phenomenon, one can
clearly distinguish between black holes and their
analogs\cite{Enomoto:2022mti,Giovanazzi:2004zv}. 
If black hole chaos\cite{Maldacena:2015waa} is due to the Stokes
phenomenon\cite{Morita:2019bfr}, we hope we can get closer to the
nature.
We hope that our methods discussed in this paper will be useful in other
areas of physics, especially in condensed matter physics.

\section{Acknowledgments}
The author would like to express his gratitude to Seishi Enomoto for his
cooperation in the early stages of this work.
In writing the revised version, we express our gratitude to
Satoshi Ohya for inviting us to the seminar at Nihon University and to the
participants of the seminar for clarifying points that needed
to be explained.

\end{document}